\documentclass[twoside,a4paper]{article}
\usepackage{ifpdf}
\ifpdf%
\usepackage{ae} % Almost European fonts
\usepackage{aeguill}
\usepackage[pdftex,backref=section]{hyperref}
\fi
\usepackage{amssymb}
\usepackage{amsmath}
\usepackage{xcolor}
\makeatletter
\def\mathcolor#1#{\@mathcolor{#1}}
\def\@mathcolor#1#2#3{%
  \protect\leavevmode
  \begingroup
    \color#1{#2}#3%
  \endgroup
}
\makeatother
\newtheorem{theorem}{Theorem}[section]
\newtheorem{lemma}{Lemma}[section]
\newtheorem{notation}{Notation}[section]
\newtheorem{notations}[notation]{Notations}
\newtheorem{definition}{Definition}[section]
\newtheorem{definitions}[definition]{Definitions}
\newtheorem{remark}{Remark}[section]
\newtheorem{remarks}[remark]{Remarks}
\newtheorem{example}{Example}[section]
\def\Group#1{\mathsf{#1}}
\def\Isotropy#1{\mathsf{#1}}
\def\matrixsize#1#2{{{#1}\times{#2}}}
\def\matrixprojectionoperator#1#2{{\textup{Id}^{{#2}}_{\matrixsize{#1}{#1}}}}

\def\MatrixProjection#1#2#3{{\overline{{#3}^{#1#2}}}}
\def\TensorProjection#1#2#3#4{{\overline{#4}^{#1#2#3}}}
\def\MatrixZero#1#2#3{{\widetilde{#3}^{#1#2}}}
\def\TensorZero#1#2#3#4{{\widetilde{#4}^{#1#2#3}}}
\def\MatrixLift#1#2#3{{\underline{#3}_{#1#2}}}
\def\TensorLift#1#2#3#4{{\underline{#4}_{#1#2#3}}}
\def\tensorproduct{\otimes}
\def\tensor#1{{\mathcal{#1}}}
\def\matrixrank#1{{\textup{rank}\,{#1}}}
\def\Trace{\textup{Trace}}
\def\Transpose#1{{{}^{t}#1}}

\def\IsotropyAction#1#2{{{#1}\diamond{#2}}}
\def\IsotropyGroupAction#1#2{{{#1}\diamond{#2}}}
\def\IdMat#1{{\textup{Id}_{\matrixsize{#1}{#1}}}}
\def\repeated#1#2{#1|#2}
\def\fixedpointcolor{cyan}
\def\accrochecolor{blue}
\newcommand{\mathsc}[1]{{\normalfont\textsc{#1}}}
\def\LadermanTensor{\tensor{L}}
\title{Laderman matrix multiplication algorithm can be constructed
  using  Strassen algorithm and related tensor's isotropies}
\author{Alexandre.Sedoglavic@univ-lille.fr}
\begin{document}
\maketitle
\section{Introduction}
In~\cite{strassen:1969}, V.\ Strassen presented a noncommutative
algorithm for multiplication of two~$\matrixsize{2}{2}$ matrices using
only~$7$ multiplications.  The current upper bound~$23$
for~$\matrixsize{3}{3}$ matrix multiplication was reached by J.B.\
Laderman in~\cite{laderman:1976a}.  This note presents a
\emph{geometric} relationship between Strassen and Laderman
algorithms. By doing so, we retrieve a \emph{geometric} formulation of
results very similar to those presented by O.\ S\'ykora
in~\cite{sykora:1977a}.
\subsection{Disclaimer: there is no improvement in this
  note}\label{sec:disclaimer}
We do not improve any practical algorithm or prove any theoretical
bound in this short note but focus on effective manipulation of tensor
associated to matrix multiplication algorithm. To do so, we present
only the minimal number of needed definitions and thus leave many
facts outside our scope. We refer to~\cite{landsberg:2010} for a
complete description of the field and to~\cite{Ambainis:2014aa} for a
state-of-the-art presentation of theoretical complexity issues.
\subsection{So, why writing (or reading) it?}
We follow the geometric spirit
of~\cite{Grochow:2016aa,Chiantini:2016aa,Burgisser:2015aa,burichenko:2015,burichenko:2014}
and related papers: symmetries could be used in practical design of
matrix multiplication algorithms. Hence, this note presents another
example of this philosophy by giving a precise geometric meaning to
the following statement:
\begin{quote}
  Laderman matrix multiplication algorithm is composed by
  four~$\matrixsize{2}{2}$ optimal matrix multiplication algorithms, a
  half of the classical~$\matrixsize{2}{2}$ matrix multiplication
  algorithm and a correction term.
\end{quote}
\section{Framework}
To do so, we have to present a small part of the classical framework
(for a complete presentation
see~{\cite{groot:1978a,groot:1978,landsberg:2010}}) mainly because we
do not take it literally and only use a simplified version.  Let us
start by some basic definitions and notations as the following generic
matrices: \small%
\begin{equation}
  \label{eq:1}
  A=\!\left(%
    \begin{array}{ccc}
      {a_{11}}&{a_{12}}&{a_{13}}\\
      {a_{21}}&{a_{22}}&{a_{23}}\\
      {a_{31}}&{a_{32}}&{a_{33}}
    \end{array}\right) \!, \
  B=\!\left(%
    \begin{array}{ccc}
      {b_{11}}&{b_{12}}&{b_{13}}\\
      {b_{21}}&{b_{22}}&{b_{23}}\\
      {b_{31}}&{b_{32}}&{b_{33}}
    \end{array}\right)\!,\
  C=\!\left(%
    \begin{array}{ccc}
      {c_{11}}&{c_{12}}&{c_{13}}\\
      {c_{21}}&{c_{22}}&{c_{23}}\\
      {c_{31}}&{c_{32}}&{c_{33}}
    \end{array}\right)\!,
\end{equation}
\normalsize%
that will be used in the sequel. Furthermore, as we also consider
their~$\matrixsize{2}{2}$ submatrices, let us introduce some
associated notations.
\begin{notations}\label{def:MatrixProjection}
  Let~$n,i,j$ be positive integers such that~${i\leq n}$
  and~${j\leq n}$.  We denote by~$\matrixprojectionoperator{n}{j}$ the
  identity~$\matrixsize{n}{n}$ matrix where the~$j$th diagonal term
  is~$0$. Given a~$\matrixsize{n}{n}$ matrix~$A$, we denote
  by~$\MatrixZero{j}{k}{A}$ the
  matrix~${\matrixprojectionoperator{n}{j}\cdot A\cdot
    \matrixprojectionoperator{n}{k}}$.  For example, the
  matrix~$\MatrixZero{3}{3}{A}$,~$\MatrixZero{3}{2}{B}$
  and~$\MatrixZero{2}{3}{C}$ are:
  \begin{equation}
    \label{eq:3}
    \left(%
      \begin{array}{ccc}
        {a_{11}}&{a_{12}}&{0}\\
        {a_{21}}&{a_{22}}&{0}\\
        {0}&{0}&{0}
      \end{array}\right)\!,
    \quad
    \left(%
      \begin{array}{ccc}
        {b_{11}}&{0}&{b_{13}}\\
        {b_{21}}&{0}&{b_{23}}\\
        {0}&{0}&{0}
      \end{array}\right)
    \quad \textup{and}\quad
    \left(%
      \begin{array}{ccc}
        {c_{11}}&{c_{12}}&{0}\\
        {0}&{0}&{0}\\
        {c_{31}}&{c_{32}}&{0}
      \end{array}\right)\!.
  \end{equation}
  Given a\,~$\matrixsize{n}{n}$ matrix~$A$, we sometimes
  consider~$\MatrixZero{i}{j}{A}$ as the~$\matrixsize{(n-1)}{(n-1)}$
  matrix~$\MatrixProjection{i}{j}{A}$ where the line and column
  composed of~$0$ are removed.
  \par
  At the opposite, given any~$\matrixsize{(n-1)}{(n-1)}$ matrix~$A$,
  we denote by~$\MatrixLift{i}{j}{A}$ the~$\matrixsize{n}{n}$ matrix
  where a line and column of~$0$ were added to~$A$ in order to
  have~${\MatrixProjection{i}{j}{\MatrixLift{i}{j}{A}}=A}$.
\end{notations}
\subsection{Strassen multiplication algorithm}
Considered as~$\matrixsize{2}{2}$ matrices, the matrix
product~${\MatrixProjection{3}{3}{C}=\MatrixProjection{3}{3}{A}\cdot\MatrixProjection{3}{3}{B}}$
could be computed using Strassen algorithm (see~\cite{strassen:1969})
by performing the following computations:
\begin{equation}
  \label{eq:StrassenMultiplicationAlgorithm}
  \begin{aligned}
    \begin{aligned}
      t_{1} &= (a_{11} + a_{22}) (b_{11} + b_{22}),   &t_{2} & = (a_{12} - a_{22})(b_{21} + b_{22}), \\
      t_{3} &= (-a_{11} + a_{21}) (b_{11} + b_{12}), &t_{4} &
      =(a_{11}+a_{12})b_{22},
    \end{aligned} \\
    \begin{aligned}
      t_{5} = a_{11} (b_{12} - b_{22}),\ t_{6} = a_{22} (-b_{11} +
      b_{21}),\ t_{7} = (a_{21} + a_{22}) b_{11},
    \end{aligned}\\
    \begin{aligned}
      c_{11} &= t_{1} + t_{2} - t_{4} + t_{6},  & c_{12} &= t_{6} + t_{7}, \\
      c_{21} &= t_{4} + t_{5}, &c_{22} &= t_{1} + t_{3} + t_{5}
      -t_{7}.
    \end{aligned}
  \end{aligned}
\end{equation}
In order to consider above algorithm under a geometric standpoint, it
is usually presented as a tensor.
\subsection{Bilinear mappings seen as tensors and associated trilinear
  forms}
\begin{definitions}\label{def:tensor}
  Given a tensor~$\tensor{T}$ decomposable as sum of rank-one tensors:
  \begin{equation}
    \label{eq:5}
    \tensor{T}=\sum_{i=1}^{r} T_{i1}\tensorproduct T_{i2}\tensorproduct T_{i3},
  \end{equation}
  where~$T_{ij}$ are~$\matrixsize{n}{n}$ matrices:
  \begin{itemize}
  \item the integer~$r$ is the \emph{tensor rank} of
    tensor~$\tensor{T}$;
  \item the unordered
    list~${[{(\matrixrank{M_{ij}})}_{j=1\ldots 3}]}_{i=1\ldots r}$ is
    called the \emph{type} of tensor~$\tensor{T}$ ($\matrixrank{A}$
    being the classical rank of the matrix~$A$).
  \end{itemize}
\end{definitions}
\subsection{Tensors' contractions}
To explicit the relationship between what is done in the sequel and
the bilinear mapping associated to matrix multiplication, let us
consider the following tensor's contractions:
\begin{definitions}\label{def:contractions}
  Using the notation of definition~\ref{def:tensor} given a
  tensor~$\tensor{T}$ and three~$\matrixsize{n}{n}$ matrices~$A,B$
  and~$C$ with coefficients in the algebra~$\mathbb{K}$:
  \begin{itemize}
  \item the~$(1,2)$ contraction
    of~$\tensor{T}\tensorproduct A\tensorproduct B$ defined by:
    \begin{equation}
      \label{eq:6}
      \sum_{i=1}^{r} \Trace (\Transpose{T_{i1}} \cdot A)\, \Trace
      (\Transpose{T_{i2}} \cdot B) T_{i3}
    \end{equation}
    corresponds to a bilinear
    application~$\mathbb{K}^{\matrixsize{n}{n}}\times
    \mathbb{K}^{\matrixsize{n}{n}} \mapsto
    \mathbb{K}^{\matrixsize{n}{n}}$ with indeterminates~$A$ and~$B$.
  \item the~$(1,2,3)$ (a.k.a.\ full) contraction
    of~$\tensor{T}\tensorproduct A\tensorproduct B\tensorproduct C$
    defined by:
    \begin{equation}
      \label{eq:7a}
      {\left\langle\tensor{T} | A
          \tensorproduct B \tensorproduct C \right\rangle} = 
      \sum_{i=1}^{r} \Trace (\Transpose{T_{i1}} \cdot A)\, \Trace
      (\Transpose{T_{i2}} \cdot B)\, \Trace(\Transpose{T_{i3}}\cdot C)
    \end{equation} 
    corresponds to a trilinear
    form~$\mathbb{K}^{\matrixsize{n}{n}}\times
    \mathbb{K}^{\matrixsize{n}{n}} \times
    \mathbb{K}^{\matrixsize{n}{n}} \mapsto \mathbb{K}$ with
    indeterminates~$A,B$ and~$C$.
  \end{itemize}
\end{definitions}
\begin{remarks}
  As the studied object is the tensor, its expressions as full or
  incomplete contractions are equivalent. Thus, even if matrix
  multiplication is a bilinear application, we are going to work in
  the sequel with trilinear forms (see~\cite{Dumas:2016aa} for
  bibliographic references on this standpoint).
  \par
  The definition in~\ref{def:contractions} are taken to express the
  full contraction as a degenerate inner product between tensors; it
  is not the usual choice made in the literature and so, we have to
  explicitly recall some notions used in the sequel.
\end{remarks}
Strassen multiplication
algorithm~(\ref{eq:StrassenMultiplicationAlgorithm}) is equivalent to
the tensor~$\tensor{S}$ defined by:\par
\footnotesize%
\begin{equation}
  \label{eq:4}
  \begin{aligned}
    & \left(\!
      \begin{array}{cc}
        1&0\\ 
        0&1\\ 
      \end{array}
      \!\right) \tensorproduct{} \left(\!
      \begin{array}{cc}
        1&0\\ 
        0&1\\ 
      \end{array}
      \!\right) \tensorproduct{} \left(\!
      \begin{array}{cc}
        1&0\\ 
        0&1\\ 
      \end{array}
      \!\right) &+ \left(\!
      \begin{array}{cc}
        0&1\\ 
        0&-1\\ 
      \end{array}
      \!\right) \tensorproduct{} \left(\!
      \begin{array}{cc}
        0&0\\ 
        1&1\\ 
      \end{array}
      \!\right) \tensorproduct{} \left(\!
      \begin{array}{cc}
        1&0\\ 
        0&0\\ 
      \end{array}
      \!\right)
    + \\[\smallskipamount]
    & \left(\!
      \begin{array}{cc}
        -1&0\\ 
        1&0\\ 
      \end{array}
      \!\right) \tensorproduct{} \left(\!
      \begin{array}{cc}
        1&1\\ 
        0&0\\ 
      \end{array}
      \!\right) \tensorproduct{} \left(\!
      \begin{array}{cc}
        0&0\\ 
        0&1\\ 
      \end{array}
      \!\right) &+ \left(\!
      \begin{array}{cc}
        1&1\\ 
        0&0\\ 
      \end{array}
      \!\right) \tensorproduct{} \left(\!
      \begin{array}{cc}
        0&0\\ 
        0&1\\ 
      \end{array}
      \!\right) \tensorproduct{} \left(\!
      \begin{array}{cc}
        -1&0\\ 
        1&0\\ 
      \end{array}
      \!\right)
    +\\[\smallskipamount]
    & \left(\!
      \begin{array}{cc}
        1&0\\ 
        0&0\\ 
      \end{array}
      \!\right) \tensorproduct{} \left(\!
      \begin{array}{cc}
        0&1\\ 
        0&-1\\ 
      \end{array}
      \!\right) \tensorproduct{} \left(\!
      \begin{array}{cc}
        0&0\\ 
        1&1\\ 
      \end{array}
      \!\right) &+ \left(\!
      \begin{array}{cc}
        0&0\\ 
        0&1\\ 
      \end{array}
      \!\right) \tensorproduct{} \left(\!
      \begin{array}{cc}
        -1&0\\ 
        1&0\\ 
      \end{array}
      \!\right) \tensorproduct{} \left(\!
      \begin{array}{cc}
        1&1\\ 
        0&0\\ 
      \end{array}
      \!\right)
    + \\[\smallskipamount]
    & \left(\!
      \begin{array}{cc}
        0&0\\ 
        1&1\\ 
      \end{array}
      \!\right) \tensorproduct{} \left(\!
      \begin{array}{cc}
        1&0\\ 
        0&0\\ 
      \end{array}
      \!\right) \tensorproduct{} \left(\!
      \begin{array}{cc}
        0&1\\ 
        0&-1\\ 
      \end{array}
      \!\right)\!.
  \end{aligned}
\end{equation}
\normalsize%
This tensor defines the matrix multiplication
algorithm~(\ref{eq:StrassenMultiplicationAlgorithm}) and its tensor
rank is~$7$.
\subsection{$\matrixsize{2}{2}$ matrix multiplication tensors induced
  by a~$\matrixsize{3}{3}$ matrix multiplication
  tensor}\label{sec:Projection}
Given any~$\matrixsize{3}{3}$ matrix multiplication tensor, one can
define~$3^{3}$ induced~$\matrixsize{2}{2}$ matrix multiplication
tensors as shown in this section.  First, let us introduce the
following operators that generalize to tensor the
notations~\ref{def:MatrixProjection}:
\begin{definitions}\label{def:Projection}
  Using notations introduced in definition~\ref{def:MatrixProjection},
  we define:
  \begin{subequations}
    \begin{align}
      \label{eq:14a}
      \TensorZero{i}{j}{k}{A\tensorproduct{} B\tensorproduct{} C}= 
      \MatrixZero{i}{j}{A} \tensorproduct{}
      \MatrixZero{j}{k}{B} \tensorproduct{}
      \MatrixZero{k}{i}{C}, \\
      \label{eq:14b}
      \TensorProjection{i}{j}{k}{A\tensorproduct{} B\tensorproduct{} C}= 
      \MatrixProjection{i}{j}{A} \tensorproduct{}
      \MatrixProjection{j}{k}{B} \tensorproduct{}
      \MatrixProjection{k}{i}{C}, 
      \\
      \label{eq:14c}
      \TensorLift{i}{j}{k}{A\tensorproduct{} B\tensorproduct{} C}= 
      \MatrixLift{i}{j}{A} \tensorproduct{}
      \MatrixLift{j}{k}{B} \tensorproduct{}
      \MatrixLift{k}{i}{C} 
    \end{align}
  \end{subequations}
  and we extend the definitions of these operators by additivity in
  order to be applied on any tensor~$\tensor{T}$ described in
  definition~\ref{def:tensor}.
\end{definitions}
There is~$n^{3}$ such projections and given any matrix multiplication
tensor~$\tensor{M}$, the full contraction satisfying the following
trivial properties:
\begin{equation}
  \label{eq:16}
  \left\langle \tensor{M} \,\big|\,\TensorZero{i}{j}{k}{A\tensorproduct{}
      B\tensorproduct{} C} \right\rangle  =
  \left\langle \TensorZero{i}{j}{k}{\tensor{M}}
    \,\big|\,A\tensorproduct{} B \tensorproduct{} C \right\rangle
  =
  \left\langle \TensorProjection{i}{j}{k}{\tensor{M}}\, \big|\, \TensorProjection{i}{j}{k}{A\tensorproduct{}
      B\tensorproduct{} C} \right\rangle 
\end{equation}
(where the projection operator apply on an~$\matrixsize{n}{n}$ matrix
multiplication tensor); it defines explicitly
a~$\matrixsize{(n-1)}{(n-1)}$ matrix multiplication tensor.
\par
The following property holds:
\begin{lemma}
  \begin{equation}
    \label{eq:15}
    {(n-1)}^{3}\left\langle\tensor{M} | A \tensorproduct{}
      B \tensorproduct{} C \right\rangle = 
    \sum_{1\leq i,j,k \leq n} 
    \left\langle \tensor{M} \Big| \TensorZero{i}{j}{k}{A\tensorproduct{}
        B\tensorproduct{} C}\right\rangle
  \end{equation}
  and thus, we have:
  \begin{equation}
    \label{eq:16a}
    \left\langle \tensor{M} | A \tensorproduct{}
      B \tensorproduct{} C \right\rangle 
    = \left\langle
      \frac{1}{{(n-1)}^{3}}\sum_{1\leq i,j,k \leq n} 
      \TensorZero{i}{j}{k}{\tensor{M}}\Big|
      A \tensorproduct{}
      B \tensorproduct{} C 
    \right\rangle\!.
  \end{equation}
\end{lemma}
The obvious facts made in this section underline the relationships
between any~$\matrixsize{n}{n}$ matrix multiplication tensor and
the~$n^{3}$ induced~$\matrixsize{(n-1)}{(n-1)}$ algorithms.
\par
Considering the Laderman matrix multiplication tensor, we are going to
explore further this kind of relationships.  First, let us introduce
this tensor.
\subsection{Laderman matrix multiplication tensor}
The Laderman tensor~$\LadermanTensor$ described below by giving its
full contraction:
\begin{equation}
  \label{eq:17}
  \begin{array}{lc}
    \left( { a_{11}}-{ a_{21}}+{ a_{12}}-{ a_{22}}-{ a_{32}}+{ a_{13}}-{ a_{33}} \right) { b_{22}}
    \,{ c_{21}} &+ \\{ a_{22}}\, \left( -{ b_{11}}+{ b_{21}}-{ b_{31}}+{ b_{12}}-{ b_{22}}-{ b_{23}}+{
    b_{33}} \right) { c_{12}} &+\\
    { a_{13}}\,{ b_{31}}\, \left( { c_{11}}+{ c_{21}}+{ c_{31}}+{ c_{12}
    }+{ c_{32}}+{ c_{13}}+{ c_{23}} \right) &+ \\
    \left( { a_{11}}-{ a_{31}}+{ a_{12}}-{ a_{22}}-{ 
    a_{32}}+{ a_{13}}-{ a_{23}} \right) { b_{23}}\,{ c_{31}}&+\\{ a_{32}}\, \left( -{ b_{11}}+{ b_{21}}-{
    b_{31}}-{ b_{22}}+{ b_{32}}+{ b_{13}}-{ b_{23}} \right) {
    c_{13}} & + \\
    { a_{11}}\,{ b_{11}}\,
    \left( { c_{11}}+{ c_{21}}+{ c_{31}}+{ c_{12}}+{ c_{22}}+{ c_{13}}+{
    c_{33}} \right) &+ \\ \left( 
    -{ a_{11}}+{ a_{31}}+{ a_{32}} \right)  \left( { b_{11}}-{ b_{13}}+{ b_{23}} \right)  \left( {
    c_{31}}+{ c_{13}}+{ c_{33}} \right) &+ \\
    \left( { a_{22}}-{ a_{13}}+{ a_{23}} \right)  \left( {
    b_{31}}+{ b_{23}}-{ b_{33}} \right)  \left( { c_{31}}+{
    c_{12}}+{ c_{32}} \right) & + \\
    \left( -{
    a_{11}}+{ a_{21}}+{ a_{22}} \right)  \left( { b_{11}}-{ b_{12}}+{ b_{22}} \right)  \left( {
    c_{21}}+{ c_{12}}+{ c_{22}} \right) & + \\ 
    \left( { a_{32}}-{ a_{13}}+{ a_{33}} \right)  \left( {
    b_{31}}+{ b_{22}}-{ b_{32}} \right)  \left( { c_{21}}+{
    c_{13}}+{ c_{23}} \right) &+\\
    \left( {
    a_{21}}+{ a_{22}} \right)  \left( -{ b_{11}}+{ b_{12}} \right)  \left( { c_{21}}+{ c_{22}}
    \right)& + \\
    \left( { a_{31}}+{ a_{32}} \right)  \left( -{ b_{11}}+{ b_{13}} \right)  \left( {
    c_{31}}+{ c_{33}} \right) & + \\
    \left( { a_{13}}-{ a_{33}} \right)  \left( { b_{22}}-{ b_{32}}
    \right)  \left( { c_{13}}+{ c_{23}} \right) &+\\
    \left( { a_{11}}-{ a_{21}} \right)  \left( -{
    b_{12}}+{ b_{22}} \right)  \left( { c_{12}}+{ c_{22}} \right)
                &+\\
    \left( { a_{32}}+{ a_{33}}
    \right)  \left( -{ b_{31}}+{ b_{32}} \right)  \left( { c_{21}}+{
    c_{23}} \right)&+\\
    \left( -{
    a_{11}}+{ a_{31}} \right)  \left( { b_{13}}-{ b_{23}} \right)  \left( { c_{13}}+{ c_{33}}
    \right) &+\\ \left( { a_{13}}-{ a_{23}} \right)  \left( { b_{23}}-{ b_{33}} \right)  \left( { 
    c_{12}}+{ c_{32}} \right) &+\\
    \left( { a_{22}}+{ a_{23}} \right)  \left( -{ b_{31}}+{ b_{33}}
    \right)  \left( { c_{31}}+{ c_{32}} \right) &+\\
    { a_{12}}\,{ b_{21}}\,{ c_{11}}+{ a_{23}}\,{ 
    b_{32}}\,{ c_{22}} +  { a_{21}}\,{ b_{13}}\,{ c_{32}}+{ a_{31}}\,{ b_{12}}\,{ c_{23}}+{ a_{33}}\,{
    b_{33}}\,{ c_{33}}
  \end{array}
\end{equation}
and was introduced in~\cite{laderman:1976a} (we do not study in this
note any other \emph{inequivalent} algorithm of same tensor rank
e.g.~\cite{johnson:1986a,courtois:2011,oh:2013a,smirnov:2013a}, etc). Considering
the projections introduced in definition~\ref{def:Projection}, we
notice that:
\begin{remark}
  Considering definitions introduced in Section~\ref{sec:Projection},
  we notice that Laderman matrix multiplication tensor defines~$4$
  optimal~$\matrixsize{2}{2}$ matrix multiplication
  tensors~$\TensorProjection{i}{j}{k}{\LadermanTensor}$
  with~${(i,j,k)}$
  in~${\lbrace (2,1,3),(2,3,2), (3,1,2),(3,3,3)\rbrace}$ and~$23$
  other with tensor rank~$8$.
\end{remark}
Further computations show that:
\begin{remark}
  The type of the Laderman matrix multiplication tensor is
  \begin{equation}
    \label{eq:18}
    \big[ \repeated{(2,2,2)}{4}, \repeated{((1,3,1), (3,1,1), (1,1,3))}{2}, \repeated{(1,1,1)}{13} \big]
  \end{equation}
  where~$\repeated{m}{n}$ indicates that~$m$ is repeated~$n$ times.
\end{remark}
\subsection{Tensors' isotropies}
We refer to~\cite{groot:1978a, groot:1978} for a complete presentation
of automorphism group operating on varieties defined by algorithms for
computation of bilinear mappings and as a reference for the following
theorem:
\begin{theorem}
  The isotropy group of the~$\matrixsize{n}{n}$ matrix multiplication
  tensor is
  \begin{equation}
    \label{eq:2}
    {\mathsc{pgl}({\mathbb{C}}^{n})}^{\times 3} \rtimes \mathfrak{S}_{3},
  \end{equation}
  where~$\mathsc{pgl}$ stands for the projective linear group
  and~$\mathfrak{S}_{3}$ for the symmetric group on~$3$ elements.
\end{theorem}
Even if we do not completely explicit the concrete action of this
isotropy group on matrix multiplication tensor, let us precise some
terminologies:
\begin{definitions}
  Given a tensor defining matrix multiplication computations, the
  orbit of this tensor is called the \emph{multiplication algorithm}
  and any of the points composing this orbit is a \emph{variant} of
  this algorithm.
\end{definitions}
\begin{remark}
  As shown in~\cite{Gesmundo:2016aa}, matrix multiplication is
  characterised by its isotropy group.
\end{remark}
\begin{remark}
  In this note, we only need
  the~${\mathsc{pgl}({\mathbb{C}}^{n})}^{\times 3}$ part of this group
  (a.k.a.\ sandwiching) and thus focus on it in the sequel.
\end{remark}
As our framework and notations differ slightly from the framework
classically found in the literature, we have to explicitly define
several well-known notions for the sake of clarity. Hence, let us
recall the \emph{sandwiching} action:
\begin{definition}
  Given~${\Isotropy{g}={(G_{1}\times G_{2} \times G_{3})}}$ an element
  of~${\mathsc{pgl}({\mathbb{C}}^{n})}^{\times 3}$, its action on a
  tensor~$\tensor{T}$ is given by:
  \begin{equation}
    \label{eq:7}
    \begin{aligned}
      \IsotropyAction{\Isotropy{g}}{\tensor{T}} &= \sum_{i=1}^{r}
      \IsotropyAction{\Isotropy{g}}{(T_{i1}\tensorproduct{} T_{i2}\tensorproduct{} T_{i3})}, \\
      \IsotropyAction{\Isotropy{g}}{(T_{i1}\tensorproduct{}
        T_{i2}\tensorproduct{} T_{i3})} &= \left(
        \Transpose{G_{1}^{-1}} T_{i1}\Transpose{G_{2}} \right)
      \tensorproduct{} \left( \Transpose{G_{2}^{-1}}
        T_{i2}\Transpose{G_{3}} \right) \tensorproduct{} \left(
        \Transpose{G_{3}^{-1}} T_{i3}\Transpose{G_{1}} \right)\!.
    \end{aligned}
  \end{equation}
\end{definition}
\begin{example}
  Let us consider the action of the following isotropy
  \begin{equation}
    \label{eq:8}
    \left(%
      \begin{array}{cc}
        0 & 1/\lambda \\
        -1 & 0 
      \end{array}\right)
    \times\left(%
      \begin{array}{cc}
        1/\lambda & -1/\lambda \\
        0 & 1
      \end{array}\right)
    \times\left(%
      \begin{array}{cc}
        -1/\lambda & 0 \\
        1 & -1 
      \end{array}\right)
  \end{equation}
  on the Strassen variant of the Strassen algorithm.  The resulting
  tensor~$\tensor{W}$ is: \par
  \scriptsize%
  \begin{equation}
    \label{eq:10}
    \begin{aligned}
      \sum_{i=1}^{7} w_{i} &= \left(\!\!
        \begin{array}{cc}
          -1&\lambda\\ 
          -\frac{1}{\lambda}&0\\ 
        \end{array}
        \!\!\right) \!\tensorproduct\! \left(\!\!
        \begin{array}{cc}
          1&-\lambda\\ 
          \frac{1}{\lambda}&0\\ 
        \end{array}
        \!\!\right) \!\tensorproduct\! \left(\!\!
        \begin{array}{cc}
          1&-\lambda\\ 
          \frac{1}{\lambda}&0\\ 
        \end{array}
        \!\!\right) + \left(\!\!
        \begin{array}{cc}
          -1&l\\ 
          -\frac{1}{\lambda}&1\\ 
        \end{array}
        \!\!\right) \!\tensorproduct\! \left(\!\!
        \begin{array}{cc}
          0&0\\ 
          1&0\\ 
        \end{array}
        \!\!\right) \!\tensorproduct\! \left(\!\!
        \begin{array}{cc}
          0&1\\ 
          0&0\\ 
        \end{array}
        \!\!\right)
      \\[\smallskipamount]
      & + \left(\!\!
        \begin{array}{cc}
          1&0\\ 
          \frac{1}{\lambda}&0\\ 
        \end{array}
        \!\!\right) \!\tensorproduct\! \left(\!\!
        \begin{array}{cc}
          1&0\\ 
          \frac{1}{\lambda}&0\\ 
        \end{array}
        \!\!\right) \!\tensorproduct\! \left(\!\!
        \begin{array}{cc}
          1&0\\ 
          \frac{1}{\lambda}&0\\ 
        \end{array}
        \!\!\right) + \left(\!\!
        \begin{array}{cc}
          0&0\\ 
          0&1\\ 
        \end{array}
        \!\!\right) \!\tensorproduct\! \left(\!\!
        \begin{array}{cc}
          0&0\\ 
          0&1\\ 
        \end{array}
        \!\!\right) \!\tensorproduct\! \left(\!\!
        \begin{array}{cc}
          0&0\\ 
          0&1\\ 
        \end{array}
        \!\!\right)
      \\[\smallskipamount]
      & + \left(\!\!
        \begin{array}{cc}
          0&0\\ 
          1&0\\ 
        \end{array}
        \!\!\right) \!\tensorproduct\! \left(\!\!
        \begin{array}{cc}
          0&1\\ 
          0&0\\ 
        \end{array}
        \!\!\right) \!\tensorproduct\! \left(\!\!
        \begin{array}{cc}
          -1&\lambda\\ 
          -\frac{1}{\lambda}&1\\ 
        \end{array}
        \!\!\right) + \left(\!\!
        \begin{array}{cc}
          1&-\lambda\\ 
          0&0\\ 
        \end{array}
        \!\!\right) \!\tensorproduct\! \left(\!\!
        \begin{array}{cc}
          1&-\lambda\\ 
          0&0\\ 
        \end{array}
        \!\!\right) \!\tensorproduct\! \left(\!\!
        \begin{array}{cc}
          1&-\lambda\\ 
          0&0\\ 
        \end{array}
        \!\!\right)
      \\[\smallskipamount]
      &+ \left(\!\!
        \begin{array}{cc}
          0&1\\ 
          0&0\\ 
        \end{array}
        \!\!\right) \!\tensorproduct\! \left(\!\!
        \begin{array}{cc}
          -1&\lambda\\ 
          -\frac{1}{\lambda}&1\\ 
        \end{array}
        \!\!\right) \!\tensorproduct\! \left(\!\!
        \begin{array}{cc}
          0&0\\ 
          1&0\\ 
        \end{array}
        \!\!\right)
    \end{aligned}
  \end{equation}
  \normalsize%
\end{example}
that is the well-known Winograd variant of Strassen algorithm.
\begin{remarks}
  We keep the parameter~$\lambda$ useless in our presentation as a
  tribute to the construction made in~\cite{chatelin:1986a} that gives
  an elegant and elementary (i.e.\ based on matrix eigenvalues)
  construction of Winograd variant of Strassen matrix multiplication
  algorithm.
  \par
  This variant is remarkable in its own as shown
  in~\cite{bshouty:1995a} because it is optimal w.r.t.\ multiplicative
  \emph{and} additive complexity.
\end{remarks}
\begin{remark}
  Tensor's type is an invariant of isotropy's action.  Hence, two
  tensors in the same orbit share the same type. Or equivalently, two
  tensors with the same type are two variants that represent the same
  matrix multiplication algorithm.
\end{remark}
This remark will allow us in Section~\ref{sec:ResultingTensor} to
recognise the tensor constructed below as a variant of the Laderman
matrix multiplication algorithm.
\section{A tensor's
  construction}\label{sec:LadermanWinogradConstruction}
Let us now present the construction of a variant of Laderman matrix
multiplication algorithm based on Winograd variant of Strassen matrix
multiplication algorithm.
\par
First, let us give the full contraction of the
tensor~$\TensorLift{1}{1}{1}{\tensor{W}}\tensorproduct{} A
\tensorproduct{} B \tensorproduct{} C$:
\begin{subequations}
  \begin{align}
    \label{eq:FULLWIN1}
    % \left\langle\tensor{W}
    %   |
    %   \Projection{1}{1}{1}(A\tensorproduct{}
    %   B\tensorproduct{}
    %   C)\right\rangle
    % =
    \left( -{ a_{22}}-{\frac {{ a_{32}}}{\lambda}}+\lambda{ a_{23}} \right)  \left( { b_{22}}+{
    \frac {{ b_{32}}}{\lambda}}-\lambda{ b_{23}} \right)  \left( { c_{22}}+{\frac {{ c_{32}}}{\lambda}}-\lambda{ c_{23}}    \right) &+ \\
    \left( { a_{22}}-\lambda{ a_{23}} \right)  \left( { b_{22}}-\lambda{ b_{23}} \right)  \left( { c_{22}}-\lambda{
    c_{23}} \right) &+ \\
    \left( { a_{22}}+{\frac {{ a_{32}}}{\lambda}} \right)  \left( { b_{22}}+{\frac {{
    b_{32}}}{\lambda}} \right)  \left( { c_{22}}+{\frac {{ c_{32}}}{\lambda}} \right) &+ \\
    \label{eq:FW1E1}\mathcolor{\accrochecolor}{{ a_{23}}\, \left( -{ b_{22}}-{\frac {{ b_{32}}}{\lambda}}+\lambda{ b_{23}}+{ b_{33}} \right) {
    c_{32}}} &+\\
    \label{eq:FW1E2}\mathcolor{\accrochecolor}{\left( -{ a_{22}}-{\frac {{ a_{32}}}{\lambda}}+\lambda{ a_{23}}+{ a_{33}} \right) { b_{32}}\,{
    c_{23}}} &+\\
    \label{eq:FW1E3}\mathcolor{\accrochecolor}{{ a_{32}}\,{ b_{23}}\, \left( -{ c_{22}}-{\frac {{ c_{32}}}{\lambda}}+\lambda{ c_{23}}+{ c_{33}}
    \right)} &+\\
    \label{eq:FP1}\mathcolor{\fixedpointcolor}{{ a_{33}}\,{ b_{33}}\,{c_{33}}} 
  \end{align}
\end{subequations}
\subsection{A Klein four-group of
  isotropies}\label{sec:KleinFourGroup}
Let us introduce now the following notations:
\begin{equation}
  \label{eq:12}
  \IdMat{3}=\left(%
    \begin{array}{ccc}
      1 & 0 & 0 \\
      0 & 1 & 0 \\
      0 & 0 & 1 
    \end{array}\right)\!
  \quad \textup{and}\quad
  P_{(12)}=\left(%
    \begin{array}{ccc}
      0 & 1 & 0 \\
      1 & 0 & 0 \\
      0 & 0 & 1 
    \end{array}\right)
\end{equation}
used to defined the following group of isotropies:
\begin{equation}
  \label{eq:KleinFourGroup}
  \Group{K} = \left\lbrace
    \begin{array}{cc}
      \Isotropy{g_{1}}={\IdMat{3}}^{\times 3}, 
      & \Isotropy{g_{2}} =\left(\IdMat{3} \times P_{(12)} \times P_{(12)}\right)\!, \\
      \Isotropy{g_{3}}=\left(P_{(12)} \times P_{(12)} \times
      \IdMat{3}\right)\!, 
      & \Isotropy{g_{4}} =\left(P_{(12)} \times \IdMat{3} \times  P_{(12)}\right)
    \end{array}
  \right\rbrace{}
\end{equation}
that is isomorphic to the Klein four-group.
\subsection{Its action on Winograd variant of Strassen algorithm}
In the sequel, we are interested in the action of Klein
four-group~(\ref{eq:KleinFourGroup}) on our Winograd variant of
Strassen algorithm:
\begin{equation}
  \label{eq:19}
  \IsotropyGroupAction{\Group{K}}{\TensorLift{1}{1}{1}{\tensor{W}}}=\sum_{\Isotropy{g}\in\Group{K}} 
  \IsotropyAction{\Isotropy{g}}{\TensorLift{1}{1}{1}{\tensor{W}}} =
  \sum_{\Isotropy{g}\in\Group{K}} \sum_{i=1}^{7}\IsotropyAction{\Isotropy{g}}{\TensorLift{1}{1}{1}{w_{i}}}
\end{equation}
As we have for any isotropy~$\Isotropy{g}$:
\begin{equation}
  \label{eq:21}
  \left\langle\IsotropyAction{\Isotropy{g}}{\TensorLift{1}{1}{1}{\tensor{W}}} | A\tensorproduct{} B
    \tensorproduct{} C\right\rangle = \left\langle \TensorLift{1}{1}{1}{\tensor{W}} | \IsotropyAction{\Isotropy{g}}{(A \tensorproduct{} B\tensorproduct{} C)}\right\rangle,
\end{equation}
the action of isotropies~$\Isotropy{g_{i}}$ is just a permutation of
our generic matrix coefficients. Hence, we have the full contraction
of the
tensor~$(\IsotropyAction{\Isotropy{g_{2}}}{\TensorLift{1}{1}{1}{\tensor{W}}})\tensorproduct{}
A \tensorproduct{} B \tensorproduct{} C$:
\begin{subequations}
  \begin{align}
    \label{eq:FULLWIN2}
    % \left\langle\tensor{W}
    %   |
    %   \Projection{1}{1}{1}(A\tensorproduct{}
    %   B\tensorproduct{}
    %   C)\right\rangle
    % =
    \left( -{ a_{21}}-{\frac {{ a_{31}}}{\lambda}}+\lambda{ a_{23}} \right)  \left( { b_{11}}+{
    \frac {{ b_{31}}}{\lambda}}-\lambda{ b_{13}} \right)  \left( { c_{12}}+{\frac {{ c_{32}}}{\lambda}}-\lambda{ c_{13}}    \right) &+ \\
    \left( { a_{21}}-\lambda{ a_{23}} \right)  \left( { b_{11}}-\lambda{ b_{13}} \right)  \left( { c_{12}}-\lambda{
    c_{13}} \right) &+ \\
    \left( { a_{21}}+{\frac {{ a_{31}}}{\lambda}} \right)  \left( { b_{11}}+{\frac {{
    b_{31}}}{\lambda}} \right)  \left( { c_{12}}+{\frac {{ c_{32}}}{\lambda}} \right) &+ \\
    \label{eq:FW2E1}\mathcolor{\accrochecolor}{{ a_{23}}\, \left( -{ b_{11}}-{\frac {{ b_{31}}}{\lambda}}+\lambda{ b_{13}}+{ b_{33}} \right) {
    c_{32}}}&+\\
    \label{eq:FW2E2}\mathcolor{\accrochecolor}{\left( -{ a_{21}}-{\frac {{ a_{31}}}{\lambda}}+\lambda{ a_{23}}+{ a_{33}} \right) { b_{31}}\,{
    c_{13}}} &+\\
    \label{eq:FW2E3}\mathcolor{\accrochecolor}{{ a_{31}}\,{ b_{13}}\, \left( -{ c_{12}}-{\frac {{ c_{32}}}{\lambda}}+\lambda{ c_{13}}+{ c_{33}}
    \right)} &+\\
    \label{eq:FP2}\mathcolor{\fixedpointcolor}{{ a_{33}}\,{ b_{33}}\,{c_{33}}},
  \end{align}
\end{subequations}
the full contraction of the
tensor~$(\IsotropyAction{\Isotropy{g_{3}}}{\TensorLift{1}{1}{1}{\tensor{W}}})\tensorproduct{}
A \tensorproduct{} B \tensorproduct{} C$:
\begin{subequations}
  \begin{align}
    \label{eq:FULLWIN3}
    % \left\langle\tensor{W}
    %   |
    %   \Projection{1}{1}{1}(A\tensorproduct{}
    %   B\tensorproduct{}
    %   C)\right\rangle
    % =
    \left( -{ a_{11}}-{\frac {{ a_{31}}}{\lambda}}+\lambda{ a_{13}} \right)  \left( { b_{12}}+{
    \frac {{ b_{32}}}{\lambda}}-\lambda{ b_{13}} \right)  \left( { c_{21}}+{\frac {{ c_{31}}}{\lambda}}-\lambda{ c_{23}}    \right) &+ \\
    \left( { a_{11}}-\lambda{ a_{13}} \right)  \left( { b_{12}}-\lambda{ b_{13}} \right)  \left( { c_{21}}-\lambda{
    c_{23}} \right) &+ \\
    \left( { a_{11}}+{\frac {{ a_{31}}}{\lambda}} \right)  \left( { b_{12}}+{\frac {{
    b_{32}}}{\lambda}} \right)  \left( { c_{21}}+{\frac {{ c_{31}}}{\lambda}} \right) &+ \\
    \label{eq:FW3E1}\mathcolor{\accrochecolor}{{ a_{13}}\, \left( -{ b_{12}}-{\frac {{ b_{32}}}{\lambda}}+\lambda{ b_{13}}+{ b_{33}} \right) {
    c_{31}}} &+\\
    \label{eq:FW3E2}\mathcolor{\accrochecolor}{\left( -{ a_{11}}-{\frac {{ a_{31}}}{\lambda}}+\lambda{ a_{13}}+{ a_{33}} \right) { b_{32}}\,{
    c_{23}}} &+\\
    \label{eq:FW3E3}\mathcolor{\accrochecolor}{{ a_{31}}\,{ b_{13}}\, \left( -{ c_{21}}-{\frac {{ c_{31}}}{\lambda}}+\lambda{ c_{23}}+{ c_{33}}
    \right)} &+\\
    \label{eq:FP3}\mathcolor{\fixedpointcolor}{{ a_{33}}\,{ b_{33}}\,{c_{33}}} 
  \end{align}
\end{subequations}
and the full contraction of the
tensor~$(\IsotropyAction{\Isotropy{g_{4}}}{\TensorLift{1}{1}{1}{\tensor{W}}})\tensorproduct{}
A \tensorproduct{} B \tensorproduct{} C$:
\begin{subequations}
  \begin{align}
    \label{eq:FULLWIN4}
    % \left\langle\tensor{W}
    %   |
    %   \Projection{1}{1}{1}(A\tensorproduct{}
    %   B\tensorproduct{}
    %   C)\right\rangle
    % =
    \left( -{ a_{12}}-{\frac {{ a_{32}}}{\lambda}}+\lambda{ a_{13}} \right)  \left( { b_{21}}+{
    \frac {{ b_{31}}}{\lambda}}-\lambda{ b_{23}} \right)  \left( { c_{11}}+{\frac {{ c_{31}}}{\lambda}}-\lambda{ c_{13}}    \right) &+ \\
    \left( { a_{12}}-\lambda{ a_{13}} \right)  \left( { b_{21}}-\lambda{ b_{23}} \right)  \left( { c_{11}}-\lambda{
    c_{13}} \right) &+ \\
    \left( { a_{12}}+{\frac {{ a_{32}}}{\lambda}} \right)  \left( { b_{21}}+{\frac {{
    b_{31}}}{\lambda}} \right)  \left( { c_{11}}+{\frac {{ c_{31}}}{\lambda}} \right) &+ \\
    \label{eq:FW4E1}\mathcolor{\accrochecolor}{{ a_{13}}\, \left( -{ b_{21}}-{\frac {{ b_{31}}}{\lambda}}+\lambda{ b_{23}}+{ b_{33}} \right) {
    c_{31}}} &+\\
    \label{eq:FW4E2}\mathcolor{\accrochecolor}{\left( -{ a_{12}}-{\frac {{ a_{32}}}{\lambda}}+\lambda{ a_{13}}+{ a_{33}} \right) { b_{31}}\,{
    c_{13}}} &+\\
    \label{eq:FW4E3}\mathcolor{\accrochecolor}{{ a_{32}}\,{ b_{23}}\, \left( -{ c_{11}}-{\frac {{ c_{31}}}{\lambda}}+\lambda{ c_{13}}+{ c_{33}}
    \right)} &+\\
    \label{eq:FP4}\mathcolor{\fixedpointcolor}{{ a_{33}}\,{ b_{33}}\,{c_{33}}}.
  \end{align}
\end{subequations}
There is several noteworthy points in theses expressions:
\begin{remarks}
  \begin{itemize}
  \item the term~(\ref{eq:FP1}) is a fixed point of~$\Group{K}$'s
    action;
  \item the trilinear terms~(\ref{eq:FW1E1}) and~(\ref{eq:FW2E1}),
    (\ref{eq:FW1E2}) and~(\ref{eq:FW3E2}), (\ref{eq:FW1E3})
    and~(\ref{eq:FW4E3}), (\ref{eq:FW2E2}) and~(\ref{eq:FW4E2}),
    (\ref{eq:FW2E3}) and~(\ref{eq:FW3E3}), (\ref{eq:FW3E1})
    and~(\ref{eq:FW4E1}) could be \emph{added} in order to obtain new
    rank-on tensors without changing the tensor rank.  For
    example~(\ref{eq:FW1E1})+(\ref{eq:FW2E1}) is equal to:
    \begin{equation}
      \label{eq:23}
      \mathcolor{\accrochecolor}{{ a_{23}}\, \left( -{ b_{22}}-{\frac
            {{ b_{32}}}{\lambda}}+
          \lambda{ b_{23}}+2 { b_{33}} -{ b_{11}}-{\frac {{
                b_{31}}}{\lambda}}+
          \lambda{ b_{13}}\right) {
          c_{32}}}.
    \end{equation}
  \end{itemize}
\end{remarks}
The tensor rank of the
tensor~${\IsotropyGroupAction{\Group{K}}{\TensorLift{1}{1}{1}{\tensor{W}}}=\sum_{\Isotropy{g}\in\Group{K}}
  \IsotropyAction{\Isotropy{g}}{\TensorLift{1}{1}{1}{\tensor{W}}}}$
is~${1+3\cdot 4 + 6=19}$.  Unfortunately, this tensor does not define
a matrix multiplication algorithm (otherwise according to the lower
bound presented in~\cite{blaser:2003}, it would be optimal and this
note would have another title and impact).
\par
In the next section, after studying the action of isotropy
group~$\Group{K}$ on the classical matrix multiplication algorithm, we
are going to show how the tensor constructed above take place in
construction of matrix multiplication tensor.
\subsection{How far are we from a multiplication tensor?}
Let us consider the classical~$\matrixsize{3}{3}$ matrix
multiplication algorithm
\begin{equation}
  \label{eq:9}
  \tensor{M} =  \sum_{1\leq i,j,k \leq 3}  e^{i}_{j} \tensorproduct{} e^{j}_{k}
  \tensorproduct{} e^{k}_{i}
\end{equation}
where~$e^{i}_{j}$ denotes the matrix with a single non-zero
coefficient~$1$ at the intersection of line~$i$ and column~$j$.  By
considering the trilinear monomial:
\begin{equation}
  \label{eq:22}
  a_{ij}b_{jk}c_{ki} =  \left\langle e^{i}_{j} \tensorproduct{} e^{j}_{k}
    \tensorproduct{} e^{k}_{i} \,\big|\, A \tensorproduct{} B \tensorproduct{} C
  \right\rangle,
\end{equation}
we describe below the action of an isotropy~$\Isotropy{g}$ on this
tensor by the induced action:
\begin{equation}
  \label{eq:11}
  \begin{aligned}
    \IsotropyAction{\Isotropy{g}}{a_{ij}b_{jk}c_{ki}}&= \left\langle
      \IsotropyAction{\Isotropy{g}}{(e^{i}_{j} \tensorproduct{}
        e^{j}_{k} \tensorproduct{} e^{k}_{i})}\,\big|\, A
      \tensorproduct{} B \tensorproduct{} C \right\rangle\!,
    \\
    &= \left\langle {e^{i}_{j} \tensorproduct{} e^{j}_{k}
        \tensorproduct{} e^{k}_{i}}\,\big|\,
      \IsotropyAction{\Isotropy{g}}{(A \tensorproduct{} B
        \tensorproduct{} C)} \right\rangle\!.
  \end{aligned}
\end{equation}
\begin{remark}
  The isotropies in~$\Group{K}$ act as a permutation on rank-one
  composant of the tensor~$\tensor{M}$: we say that the
  group~$\Group{K}$ is a \emph{stabilizer} of~$\tensor{M}$.  More
  precisely, we have the following~$9$ orbits represented by the
  trilinear monomial sums:\par
  \footnotesize%
  \begin{subequations}
    \begin{align}
      \label{eq:20:1}
      \sum_{i=1}^{4}
      \IsotropyAction{\Isotropy{g_{i}}}{{a_{11}}\,{b_{11}}\,{c_{11}}} & = 
                                                                        {a_{11}}\,{b_{11}}\,{c_{11}}
                                                                        +
                                                                        {a_{12}}\,{b_{22}}\,{c_{21}}
                                                                        +
                                                                        {a_{22}}\,{b_{21}}\,{c_{12}} 
                                                                        +
                                                                        {a_{21}}\,{b_{12}}\,{c_{22}}, 
      \\       \label{eq:20:2}
      \sum_{i=1}^{4} \IsotropyAction{\Isotropy{g_{i}}}{\mathcolor{\accrochecolor}{{a_{22}}\,{b_{22}}\,{c_{22}}}} &= 
                                                                                                                   \mathcolor{\accrochecolor}{{a_{22}}\,{b_{22}}\,{c_{22}}}
                                                                                                                   +
                                                                                                                   {a_{21}}\,{b_{11}}\,{c_{12}} 
                                                                                                                   +
                                                                                                                   {a_{11}}\,{b_{12}}\,{c_{21}}
                                                                                                                   +
                                                                                                                   {a_{12}}\,{b_{21}}\,{c_{11}}, 
      \\       \label{eq:20:3}
      \sum_{i=1}^{4}
      \IsotropyAction{\Isotropy{g_{i}}}{\mathcolor{\accrochecolor}{{a_{22}}\,{b_{23}}\,{c_{32}}}}&=
                                                                                                   \mathcolor{\accrochecolor}{{a_{22}}\,{b_{23}}\,{c_{32}}}
                                                                                                   +
                                                                                                   {a_{21}}\,{b_{13}}\,{c_{32}}
                                                                                                   +
                                                                                                   {a_{11}}\,{b_{13}}\,{c_{31}}
                                                                                                   +
                                                                                                   {a_{12}}\,{b_{23}}\,{c_{31}},
      \\       \label{eq:20:4}
      \sum_{i=1}^{4} \IsotropyAction{\Isotropy{g_{i}}}{\mathcolor{\accrochecolor}{{a_{23}}\,{b_{32}}\,{c_{22}}}}&=
                                                                                                                  \mathcolor{\accrochecolor}{{{a_{23}}\,{b_{32}}\,{c_{22}}}}
                                                                                                                  +
                                                                                                                  {a_{23}}\,{b_{31}}\,{c_{12}}
                                                                                                                  +
                                                                                                                  {a_{13}}\,{b_{32}}\,{c_{21}}
                                                                                                                  +
                                                                                                                  {a_{13}}\,{b_{31}}\,{c_{11}},
      \\       \label{eq:20:5}
      \sum_{i=1}^{4}
      \IsotropyAction{\Isotropy{g_{i}}}{\mathcolor{\accrochecolor}{{a_{32}}\,{b_{22}}\,{c_{23}}}}&=
                                                                                                   \mathcolor{\accrochecolor}{{a_{32}}\,{b_{22}}\,{c_{23}}}
                                                                                                   +
                                                                                                   {a_{31}}\,{b_{11}}\,{c_{13}}
                                                                                                   +
                                                                                                   {a_{31}}\,{b_{12}}\,{c_{23}}
                                                                                                   +
                                                                                                   {a_{32}}\,{b_{21}}\,{c_{13}},
      \\       \label{eq:20:6}
      \frac{1}{2}\sum_{i=1}^{4}
      \IsotropyAction{\Isotropy{g_{i}}}{\mathcolor{\accrochecolor}{{a_{23}}\,{b_{33}}\,{c_{32}}}} &=
                                                                                                    \mathcolor{\accrochecolor}{{a_{23}}\,{b_{33}}\,{c_{32}}}
                                                                                                    +{a_{13}}\,{b_{33}}\,{c_{31}}, 
      \\      \label{eq:20:7}
      \frac{1}{2}\sum_{i=1}^{4}
      \IsotropyAction{\Isotropy{g_{i}}}{\mathcolor{\accrochecolor}{{a_{32}}\,{b_{23}}\,{c_{33}}}}
                                                                      &=
                                                                        \mathcolor{\accrochecolor}{{a_{32}}\,{b_{23}}\,{c_{33}}}
                                                                        +{a_{31}}\,{b_{13}}\,{c_{33}},
      \\  \label{eq:20:8}
      \frac{1}{2}\sum_{i=1}^{4}
      \IsotropyAction{\Isotropy{g_{i}}}{\mathcolor{\accrochecolor}{{a_{33}}\,{b_{32}}\,{c_{23}}}}&=
                                                                                                   \mathcolor{\accrochecolor}{{a_{33}}\,{b_{32}}\,{c_{23}}}
                                                                                                   +{a_{33}}\,{b_{31}}\,{c_{13}}, 
      \\       \label{eq:20:9}
      \frac{1}{4} \sum_{i=1}^{4}
      \IsotropyAction{\Isotropy{g_{i}}}{\mathcolor{\accrochecolor}{{a_{33}}\,{b_{33}}\,{c_{33}}}}
                                                                      &=
                                                                        \mathcolor{\accrochecolor}{{a_{33}}\,{b_{33}}\,{c_{33}}}. 
    \end{align}
  \end{subequations}
  \normalsize
\end{remark}
Hence, the action of~$\Group{K}$ decomposes the classical matrix
multiplication tensor~$\tensor{M}$ as a transversal action
of~$\Group{K}$ on the implicit
projection~$\TensorZero{1}{1}{1}{\tensor{M}}$, its action on the
rank-one
tensor~${ e^{1}_{1} \tensorproduct{} e^{1}_{1} \tensorproduct{}
  e^{1}_{1}}$ and a correction term also related to orbits
under~$\Group{K}$:
\begin{equation}
  \label{eq:25b}
  \begin{aligned}
    \tensor{M}& = \IsotropyGroupAction{\Group{K}}{\left( e^{1}_{1}
        \tensorproduct{} e^{1}_{1} \tensorproduct{} e^{1}_{1}\right)}
    +
    \IsotropyGroupAction{\Group{K}}{\TensorZero{1}{1}{1}{\tensor{M}}} - \tensor{R}, \\
    \tensor{R} &=(1/2)\, \IsotropyGroupAction{\Group{K}}{\left(
        e^{2}_{3} \tensorproduct{} e^{3}_{3}\tensorproduct{}
        e^{3}_{2}\right)} +(1/2)\,
    \IsotropyGroupAction{\Group{K}}{\left( e^{3}_{3} \tensorproduct{}
        e^{3}_{2}\tensorproduct{} e^{2}_{3}\right)}\\
    &+(1/2)\, \IsotropyGroupAction{\Group{K}}{\left( e^{3}_{2}
        \tensorproduct{} e^{2}_{3}\tensorproduct{} e^{3}_{3}\right)} +
    3\,\IsotropyGroupAction{\Group{K}}{\left( e^{3}_{3}
        \tensorproduct{} e^{3}_{3}\tensorproduct{} e^{3}_{3}\right)}.
  \end{aligned}
\end{equation}
\subsection{Resulting matrix multiplication
  algorithm}\label{sec:ResultingTensor}
The term~${\TensorZero{1}{1}{1}{\tensor{M}}}$ is a~$\matrixsize{2}{2}$
matrix multiplication algorithm that could be replaced by any other
one.  Choosing~$\TensorLift{1}{1}{1}{\tensor{W}}$, we have the
following properties:
\begin{itemize}
\item the tensor rank
  of~$\IsotropyGroupAction{\Group{K}}{\TensorLift{1}{1}{1}{\tensor{W}}}$
  is~$19$;
\item its addition with the correction term~$\tensor{R}$ does not
  change its tensor rank.
\end{itemize}
Hence, we obtain a matrix multiplication tensor with
rank~${23(={19+4})}$.  Furthermore, the resulting tensor have the same
type than the Laderman matrix multiplication tensor, and thus it is a
variant of the same algorithm.
\par
We conclude that the Laderman matrix multiplication algorithm can be
constructed using the orbit of an optimal~$\matrixsize{2}{2}$ matrix
multiplication algorithm under the action of a given group leaving
invariant classical~$\matrixsize{3}{3}$ matrix multiplication
variant/algorithm and with a transversal action on one of its
projections.
\section{Concluding remarks}
All the observations presented in this short note came from an
experimental mathematical approach using the computer algebra system
Maple~\cite{monagan:2007a}.  While implementing effectively (if not
efficiently) several tools needed to manipulate matrix multiplication
tensor---tensors, their isotropies and contractions, etc.---in order
to understand the theory, the relationship between the Laderman matrix
multiplication algorithm and the Strassen algorithm became clear by
simple computations that will be tedious or impossible by hand.
\par
As already shown in~\cite{sykora:1977a}, this kind of geometric
configuration could be found and used with other matrix size.
\par
The main opinion supported by this work is that symmetries play a
central role in effective computation for matrix multiplication
algorithm and that only a geometrical interpretation may brings
further improvement.
\paragraph{Acknowledgment.}
The author would like to thank Alin Bostan for providing information
on the work~\cite{sykora:1977a}.  \bibliographystyle{acm}
%\bibliography{RFC1703.bib} \def\gathen#1{{#1}}\def\cprime{$'$}

\appendix
\section{A cyclic isotropy group of order~$4$ leading to same
  resulting tensor}
Instead of Klein-four group~$\Group{K}$ presented in
Section~\ref{sec:KleinFourGroup}, one can also use another cyclic
group~$\Group{C}$ of order~$4$ that is a stabilizer
of~$\LadermanTensor$ but such that its generator~$\Isotropy{f}$ is not
a sandwiching. We do not give any further details here to avoid
supplementary definitions but nevertheless, we present an example of
the resulting~$8$ orbits, again represented by trilinear monomial
sums:\par
\footnotesize
\begin{subequations}
  \begin{align}
    \label{eq:30:1}
    \sum_{i=1}^{4}
    \IsotropyAction{\Isotropy{f}^{i}}{{a_{11}}\,{b_{11}}\,{c_{11}}} & = 
                                                                      {a_{11}}\,{b_{11}}\,{c_{11}}
                                                                      +
                                                                      {a_{12}}\,{b_{22}}\,{c_{21}}
                                                                      +
                                                                      {a_{22}}\,{b_{21}}\,{c_{12}} 
                                                                      +
                                                                      {a_{21}}\,{b_{12}}\,{c_{22}}, 
    \\\label{eq:30:2}
    \sum_{i=1}^{4} \IsotropyAction{\Isotropy{f}^{i}}{\mathcolor{\accrochecolor}{{a_{22}}\,{b_{22}}\,{c_{22}}}} &= 
                                                                                                                 \mathcolor{\accrochecolor}{{a_{22}}\,{b_{22}}\,{c_{22}}}
                                                                                                                 +
                                                                                                                 {a_{21}}\,{b_{11}}\,{c_{12}} 
                                                                                                                 +
                                                                                                                 {a_{11}}\,{b_{12}}\,{c_{21}}
                                                                                                                 +
                                                                                                                 {a_{12}}\,{b_{21}}\,{c_{11}}, 
    \\\label{eq:30:3}
    \sum_{i=1}^{4}
    \IsotropyAction{\Isotropy{f}^{i}}{\mathcolor{\accrochecolor}{{a_{22}}\,{b_{23}}\,{c_{32}}}}&=
                                                                                                 \mathcolor{\accrochecolor}{{a_{22}}\,{b_{23}}\,{c_{32}}}
                                                                                                 +
                                                                                                 {a_{21}}\,{b_{13}}\,{c_{32}}
                                                                                                 +
                                                                                                 {a_{11}}\,{b_{13}}\,{c_{31}}
                                                                                                 +
                                                                                                 {a_{12}}\,{b_{23}}\,{c_{31}},
    \\ \label{eq:30:4}
    \sum_{i=1}^{4}
    \IsotropyAction{\Isotropy{f}^{i}}{\mathcolor{\accrochecolor}{{a_{32}}\,{b_{22}}\,{c_{23}}}}&=
                                                                                                 \mathcolor{\accrochecolor}{{a_{32}}\,{b_{22}}\,{c_{23}}}
                                                                                                 +
                                                                                                 {a_{31}}\,{b_{12}}\,{c_{23}}
                                                                                                 +
                                                                                                 {a_{23}}\,{b_{31}}\,{c_{12}}
                                                                                                 +
                                                                                                 {a_{13}}\,{b_{31}}\,{c_{11}},
    \\ \label{eq:30:5}
    \sum_{i=1}^{4} \IsotropyAction{\Isotropy{f}^{i}}{\mathcolor{\accrochecolor}{{a_{23}}\,{b_{32}}\,{c_{22}}}}&=
                                                                                                                \mathcolor{\accrochecolor}{{{a_{23}}\,{b_{32}}\,{c_{22}}}}
                                                                                                                +
                                                                                                                {a_{31}}\,{b_{11}}\,{c_{13}}
                                                                                                                +
                                                                                                                {a_{13}}\,{b_{32}}\,{c_{21}}
                                                                                                                +
                                                                                                                {a_{32}}\,{b_{21}}\,{c_{13}},
    \\ \label{eq:30:6}
    \sum_{i=1}^{4}
    \IsotropyAction{\Isotropy{f}^{i}}{\mathcolor{\accrochecolor}{{a_{32}}\,{b_{23}}\,{c_{33}}}}
                                                                    &=
                                                                      \mathcolor{\accrochecolor}{{a_{32}}\,{b_{23}}\,{c_{33}}}
                                                                      +{a_{31}}\,{b_{13}}\,{c_{33}}
                                                                      +
                                                                      \mathcolor{\accrochecolor}{{a_{23}}\,{b_{33}}\,{c_{32}}}
                                                                      +{a_{13}}\,{b_{33}}\,{c_{31}}, 
    \\ \label{eq:30:7}
    \frac{1}{2}\sum_{i=1}^{4}
    \IsotropyAction{\Isotropy{f}^{i}}{\mathcolor{\accrochecolor}{{a_{33}}\,{b_{32}}\,{c_{23}}}}&=
                                                                                                 \mathcolor{\accrochecolor}{{a_{33}}\,{b_{32}}\,{c_{23}}}
                                                                                                 +{a_{33}}\,{b_{31}}\,{c_{13}}, 
    \\ \label{eq:30:8}
    \frac{1}{4} \sum_{i=1}^{4}
    \IsotropyAction{\Isotropy{f}^{i}}{\mathcolor{\accrochecolor}{{a_{33}}\,{b_{33}}\,{c_{33}}}}
                                                                    &=
                                                                      \mathcolor{\accrochecolor}{{a_{33}}\,{b_{33}}\,{c_{33}}}. 
  \end{align}
\end{subequations}
\normalsize%
Hence, the action of~$\Group{C}$ decomposes the classical matrix
multiplication tensor~$\tensor{M}$ as a transversal action
of~$\Group{C}$ on the implicit
projection~$\TensorZero{1}{1}{1}{\tensor{M}}$, its action on the
rank-one
tensor~${ e^{1}_{1} \tensorproduct{} e^{1}_{1} \tensorproduct{}
  e^{1}_{1}}$ and a correction term also related to orbits
under~$\Group{C}$:
\begin{equation}
  \label{eq:25t}
  \begin{aligned}
    \tensor{M}& = \IsotropyGroupAction{\Group{C}}{\left( e^{1}_{1}
        \tensorproduct{} e^{1}_{1} \tensorproduct{} e^{1}_{1}\right)}
    +
    \IsotropyGroupAction{\Group{C}}{\TensorZero{1}{1}{1}{\tensor{M}}} - \tensor{R}, \\
    \tensor{R} &=(1/2)\, \IsotropyGroupAction{\Group{C}}{\left(
        e^{3}_{3} \tensorproduct{} e^{3}_{2}\tensorproduct{}
        e^{2}_{3}\right)} + 3\,\IsotropyGroupAction{\Group{C}}{\left(
        e^{3}_{3} \tensorproduct{}
        e^{3}_{3}\tensorproduct{} e^{3}_{3}\right)}\\
    &+\IsotropyGroupAction{\Group{C}}{\left( e^{3}_{2}
        \tensorproduct{} e^{2}_{3}\tensorproduct{} e^{3}_{3}\right)}.
  \end{aligned}
\end{equation}
Even if the groups~$\Group{K}$ and~$\Group{C}$ are different, the
resulting actions on the coefficients of matrices~$A, B$ and~$C$
define the same orbit (in fact, the following identity holds:
(\ref{eq:30:6})=(\ref{eq:20:6})+(\ref{eq:20:7}),
(\ref{eq:30:4})+(\ref{eq:30:5})=(\ref{eq:20:4})+(\ref{eq:20:5}) and
the other orbits are identical).  Hence, the conclusions done
in~Section~\ref{sec:ResultingTensor} remain the same.
\section{Stabilizer group of isotropies}
It is shown in~\cite{burichenko:2015} that the stabilizer group of
Laderman matrix multiplication algorithm is isomorphic
to~$\mathfrak{S}_{4}$. This group is also a stabilizer of
classical~$\matrixsize{3}{3}$ matrix multiplication
algorithm~$\tensor{M}$.  Mutatis mutandis, we have with our notations
and in the coordinates used in this note:
\begin{equation}
  \label{eq:25}
  \begin{aligned}
    6\tensor{M}& =\IsotropyGroupAction{\mathfrak{S}_{4}}{\left(
        e^{1}_{1} \tensorproduct{} e^{1}_{1} \tensorproduct{}
        e^{1}_{1}\right)} +
    \IsotropyGroupAction{\mathfrak{S}_{4}}{\TensorZero{1}{1}{1}{\tensor{M}}} - \tensor{R}, \\
    \tensor{R} &= (6/4)\,
    \IsotropyGroupAction{\mathfrak{S}_{4}}{\left( e^{3}_{3}
        \tensorproduct{} e^{3}_{2}\tensorproduct{} e^{2}_{3}\right)}
    +18\,\IsotropyGroupAction{\mathfrak{S}_{4}}{\left( e^{3}_{3}
        \tensorproduct{} e^{3}_{3}\tensorproduct{} e^{3}_{3}\right)}.
  \end{aligned}
\end{equation}
This kind of relations holds for any nontrivial subgroup
of~$\mathfrak{S}_{4}$; for example with its dihedral
subgroup~$\Group{D}_{4}$, we have:
\begin{equation}
  \label{eq:26}
  \begin{aligned}
    2\tensor{M}& =\IsotropyGroupAction{\Group{D}_{4}}{\left( e^{1}_{1}
        \tensorproduct{} e^{1}_{1} \tensorproduct{} e^{1}_{1}\right)}
    +
    \IsotropyGroupAction{\Group{D}_{4}}{\TensorZero{1}{1}{1}{\tensor{M}}} - \tensor{R}, \\
    \tensor{R} &= \IsotropyGroupAction{\Group{D}_{4}}{\left( e^{2}_{3}
        \tensorproduct{} e^{3}_{3}\tensorproduct{} e^{3}_{2}\right)} +
    (1/2)\, \IsotropyGroupAction{\Group{D}_{4}}{\left( e^{3}_{3}
        \tensorproduct{} e^{3}_{2}\tensorproduct{}
        e^{2}_{3}\right)} \\
    &+6\,\IsotropyGroupAction{\Group{D}_{4}}{\left( e^{3}_{3}
        \tensorproduct{} e^{3}_{3}\tensorproduct{} e^{3}_{3}\right)}.
  \end{aligned}
\end{equation}
\end{document}